\documentclass[prl,twocolumn,superscriptaddress]{revtex4-1}
\usepackage{graphicx}
\usepackage{amssymb}
\usepackage{gensymb}

\begin{document}
\title{Bose-Einstein condensation of magnons and spin superfluidity in the polar\\ phase of $^3$He}

\author{S.~Autti}
\affiliation{Low Temperature Laboratory, Department of Applied Physics, Aalto University, P.O. Box 15100, FI-00076 AALTO, Finland}
\author{V.\,V.~Dmitriev}
\affiliation{P.\,L.~Kapitza Institute for Physical Problems of RAS, 119334 Moscow, Russia}
\author{J.\,T.~M\"{a}kinen}
\affiliation{Low Temperature Laboratory, Department of Applied Physics, Aalto University, P.O. Box 15100, FI-00076 AALTO, Finland}
\author{J.~Rysti}
\affiliation{Low Temperature Laboratory, Department of Applied Physics, Aalto University, P.O. Box 15100, FI-00076 AALTO, Finland}
\author{A.\,A.~Soldatov}
\affiliation{P.\,L.~Kapitza Institute for Physical Problems of RAS, 119334 Moscow, Russia}
\affiliation{Moscow Institute of Physics and Technology, 141700 Dolgoprudny, Russia}
\author{G.\,E.~Volovik}
\affiliation{Low Temperature Laboratory, Department of Applied Physics, Aalto University, P.O. Box 15100, FI-00076 AALTO, Finland}
\affiliation{Landau Institute for Theoretical Physics, 142432 Chernogolovka, Russia}
\author{A.\,N.~Yudin}
\affiliation{P.\,L.~Kapitza Institute for Physical Problems of RAS, 119334 Moscow, Russia}
\author{V.\,B.~Eltsov}
\affiliation{Low Temperature Laboratory, Department of Applied Physics, Aalto University, P.O. Box 15100, FI-00076 AALTO, Finland}

\date{\today}

\begin{abstract}
The polar phase of $^3$He, which is topological spin-triplet superfluid with the Dirac nodal line in the spectrum of Bogolubov quasiparticles, has been recently stabilized in a nanoconfined geometry. We pump magnetic excitations (magnons) into the sample of polar phase and observe how they form a Bose-Einstein condensate, revealed by coherent precession of the magnetization of the sample. Spin superfluidity, which supports this coherence, is associated with the spontaneous breaking of $U(1)$ symmetry by the phase of precession. We observe the corresponding Nambu-Goldstone boson and measure its mass emerging when applied rf field violates the $U(1)$ symmetry explicitly. We suggest that the magnon BEC in the polar phase is a powerful probe for topological objects such as vortices and solitons and topological nodes in the fermionic spectrum.
\end{abstract}

\maketitle

{\it Introduction.---}The phenomenon of Bose-Einstein condensation, originally suggested for real particles and observed in ultracold gases, has been extended in recent experimental and theoretical works to systems of bosonic quasiparticles, including collective modes. Examples are longitudinal electric modes \cite{Frohlich1968}, phonons \cite{Kagan2007}, excitons \cite{Butov2001}, exciton-polaritons \cite{Kasprzak2006}, photons \cite{Klaers2010}, rotons \cite{Melnikovsky2011}, and magnons \cite{HPD1,HPD3,Fom1,Demokritov2006,Pick,Qball,AuttiSelfTrapping,BEC,Vasiliev,FMgas,Bozhko}. In these systems quasiparticles are externally pumped, but they are sufficiently long-lived, so that their number $N$ is quasi-conserved. As a result, the chemical potential $\mu = dE/dN$ is non-zero during the lifetime of the condensate.

Bose-Einstein condensate (BEC) of magnons was first discovered in the B phase of $^3$He \cite{HPD1}. In this spin-triplet superfluid, magnons are quanta of transverse spin waves, associated with precessing spin of $^3$He nuclei. Magnon condensation results in spontaneous coherence of the precession, which produces a characteristic signal in nuclear magnetic resonance (NMR) experiments \cite{HPD1}. In the experiment magnons, carrying spin $-\hbar$, are pumped using radio-frequency (rf) pulse, which deflects magnetization $\mathbf M$ (or spin $\mathbf S$) from the equilibrium direction along the magnetic field $\mathbf H\parallel\hat{\mathbf z}$. Alternatively, magnons can be continuously replenished with small rf field $\mathbf{H}_{\rm rf}\perp\mathbf{H}$ to compensate magnetic relaxation \cite{HPD3}.

The coherent precession $(S_x+iS_y)\propto e^{i(\omega t+\phi)}$ is characterized by a common frequency $\omega$ and definite phase $\phi$. Formation of the coherent phase $\phi$ across the whole sample reveals the spontaneously broken $SO(2)$ spin rotation symmetry. In the language of magnon BEC this corresponds to the breaking of the $U(1)$ symmetry which characterizes the (approximate) conservation law for the number of magnons: $N_M=\int dV\,(S-S_z)/\hbar$, while the chemical potential determines the frequency of precession $\mu=dE/dN_M=\hbar\omega$.

Spontaneous breaking of $U(1)$ symmetry related to particle number conservation is linked to the superfluid phase transition. In the case of magnon BEC this is spin superfluidity. Experiments in $^3$He-B demonstrated various phenomena which accompany the spin superfluidity, such as ac and dc Josephson effects, spin supercurrents, and phase-slip processes \cite{J87,J88,J89}. Another important marker of the spontaneous $U(1)$ symmetry breaking is appearance of Nambu-Goldstone (NG) mode (which is a phonon in a usual superfluid)\cite{Vol08}. For magnon condensates in $^3$He-B such mode was indeed experimentally found \cite{Zavj,Skyba}.

Besides demonstrating the fascinating phenomenon of spin superfluidity, magnon BEC in $^3$He-B proved to be a sensitive probe for topological structures of the order parameter, like quantized vortices and their dynamics \cite{HPD4,HPD5,coretext,frontdyn}, for fermionic quasiparticles \cite{Heik} and for bosonic collective modes \cite{LightHiggs}. This coherent probe can be made local by trapping magnons in magnetic and textural traps \cite{Pick,Qball}. For a sufficiently large number of pumped magnons, the condensate deforms the trap \cite{AuttiSelfTrapping} which leads to the formation of a self-trapped magnon BEC \cite{AuttiQball}. The latter is an exact implementation of the $Q$-balls studied in the relativistic quantum field theories, which shows that magnon BEC can also be used for quantum simulations.

\begin{figure}[t]
\includegraphics[width=\columnwidth]{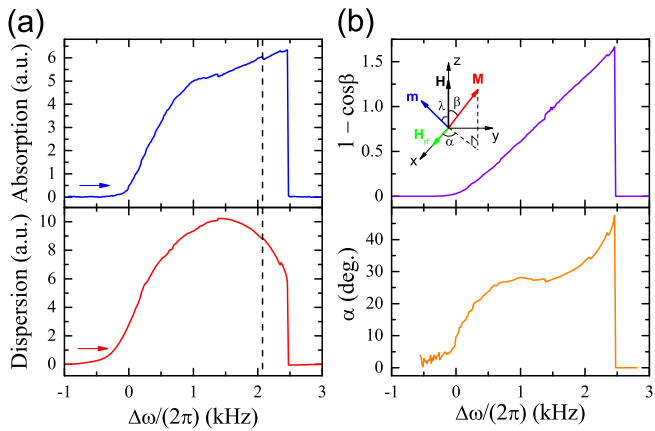}
\caption{(color online).
(a) The cw NMR signal in the polar phase of $^3$He in nafen showing creation of coherently precessing state on sweeping down magnetic field $H$ with $H_{\rm rf}=0.32$\,$\mu$T at $T=0.41T_c$, $P=6.9$\,bar, and $\lambda=90^\circ$. On the horizontal axis the frequency shift $\omega_{\rm rf} - \gamma H$ is shown. (b) Tipping angle $\beta$ and phase of precession $\alpha$ of magnetization $\bf M$ determined from the absorption and dispersion signals in panel (a). Definitions of the angles are given in the insert. Magnetization $\bf M$ is in a rotating frame of precession. Absorption and dispersion signals are proportional to $M_y=M\sin\beta\sin\alpha$ and $M_x=M\sin\beta\cos\alpha$, respectively. Relation between $\Delta\omega$ and $\cos\beta$ is linear in accordance with Eq.~(\ref{spectrum}).}
\label{BEC}
\end{figure}

All these features call for a search for magnon condensation in other topological superfluids. Coherent precession of magnetization was predicted to exist in superfluid $^3$He-A \cite{Aphase} and its observation was reported in the A-like phase in silica aerogel \cite{CPA}. Here we demonstrate magnon BEC in the recently discovered polar phase of superfluid $^3$He. We observe the coherent precession of magnetization using NMR techniques. We also measure the collective NG mode of the condensate as a function of temperature, rf excitation amplitude, precession frequency and magnetic field orientation.

{\it Polar phase.---}The polar phase is realized in liquid $^3$He confined within nafen \cite{Dmit1}, a commercially produced nanostructured material that consists of nearly parallel Al$_2$O$_3$ strands \cite{aero}. The order parameter in the polar phase is
\begin{equation}
A_{\nu j}=\Delta_0e^{i\varphi}\hat{d}_\nu\hat{m}_j,
\label{pol}
\end{equation}
where $\Delta_0$ is the gap parameter, $e^{i\varphi}$ is the phase factor, $\hat{\bf d}$ and $\hat{\bf m}$  are the unit vectors of spin and orbital anisotropy, respectively. In nafen $\hat{\bf m}$ is locked parallel to the strands \cite{AI}. The polar phase is Dirac superfluid which belongs to the same class of topological matter as Dirac nodal-line semimetals \cite{Dirac1,Dirac2,Dirac3}. As distinct from the fully gapped $^3$He-B and from $^3$He-A with Weyl nodes, the gap in the polar phase has a line of zeros in the plane normal to $\hat{\bf m}$.

\begin{figure}[t]
\includegraphics[width=\columnwidth]{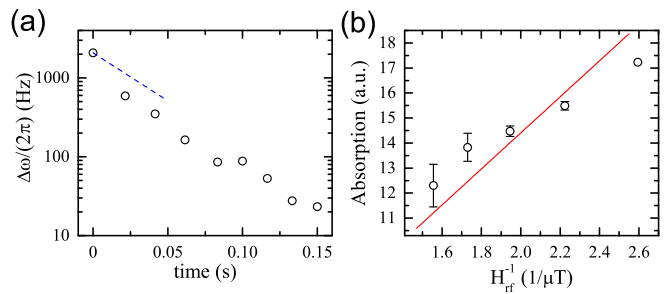}
\caption{(color online).
(a) The frequency shift in a free induction decay signal recorded after turning off the rf excitation at the point marked by a dashed line in Fig.~\ref{BEC}a. Frequency is obtained using sliding fast Fourier transform of a raw signal with a time window of 20\,ms. Dashed curve shows the expected initial time dependence of $\Delta\omega$ calculated from cw NMR data in Fig.~\ref{BEC}a. (b) Cw NMR absorption from magnon BEC versus $H_{\rm rf}^{-1}$ at the fixed $\Delta\omega/\left(2\pi\right)=435$\,Hz at $T=0.43T_c$, $P=7.1$\,bar, and $\lambda=90^\circ$. Solid line is a linear fit through zero.}
\label{lines}
\end{figure}

\begin{figure*}[t]
\includegraphics[width=\textwidth]{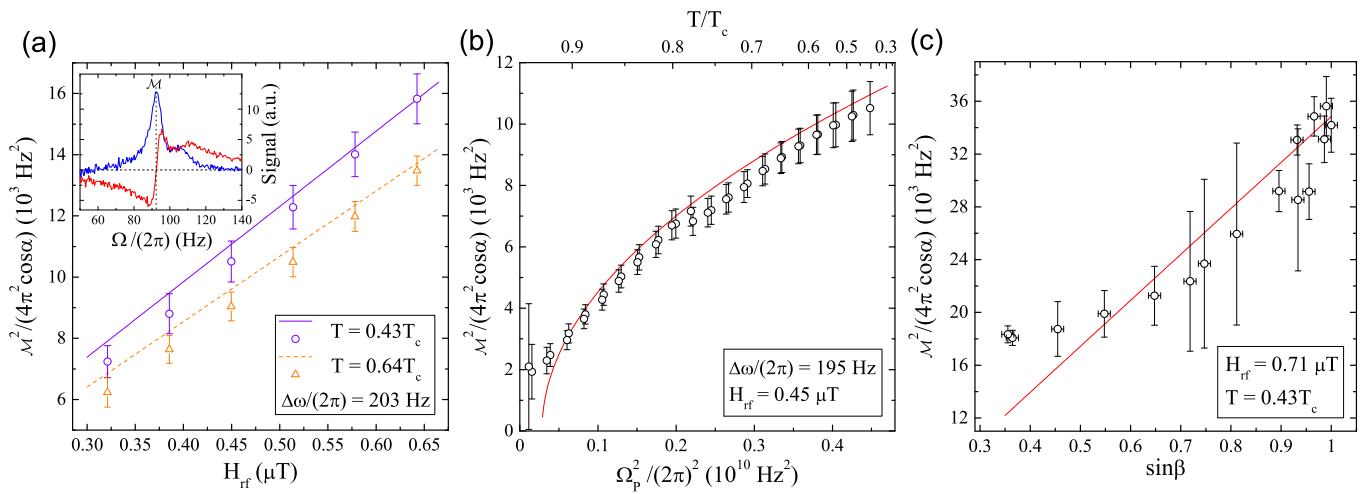}
\caption{(color online).
The mass $\mathcal{M}$ of the pseudo-Nambu-Goldstone mode in magnon BEC supported by cw NMR in the polar phase of $^3$He as a function of $H_{\rm rf}$ (a), $\Omega_P$ (b), and $\sin\beta$ (c) at $P=7.1$\,bar and $\lambda=90^\circ$. Symbols are experimental data, curves are theoretical predictions of Eq.~(\ref{mass}) without fitting. Inset to panel (a) shows an example of excitation spectrum of magnon BEC measured as described in the text. $\mathcal M$ is given by the frequency of the largest peak, while the error bar is determined as the peak width. The Leggett frequency $\Omega_P(T)$ is determined from cw NMR spectra at $\lambda=0$. Values of $\cos\alpha$, which depend on $H_{\rm rf}$ and $\Delta\omega$, are calculated from absorption and dispersion signals, while $\sin\beta$ is determined from Eq.~(\ref{spectrum}). }
\label{mode}
\end{figure*}

{\it Experiment.---}The nafen sample is a cube with a side of 4\,mm. It has porosity of 94\% and density of 0.243\,g/cm$^3$. The strands are of diameter 9\,nm, separated on average by 35\,nm \cite{aero}. Experiments are performed at pressures 6.9--7.1\,bar using pulsed and continuous-wave (cw) NMR in magnetic field of 11.2\,mT, corresponding to the NMR frequency of 362.8\,kHz. The static magnetic field $\bf H$ can be applied at an arbitrary angle $\lambda$ with respect to $\hat{\mathbf m}$. The sample is cooled down in the ROTA nuclear demagnetization refrigerator \cite{ROTA} and the temperature is measured by a quartz tuning fork \cite{fork}. The fork is calibrated against the NMR spectra measured in the linear regime using known Leggett frequency in bulk $^3$He-B \cite{3heB1,3heB2} and in the polar phase \cite{Dmit1,HQV}. To avoid formation of paramagnetic solid $^3$He on the surfaces, the sample is preplated by about 2.5 atomic layers of $^4$He. The magnitude of the rf magnetic field $H_{\rm rf}\ll H$ is calibrated with a $\pi/2$ NMR pulse in normal $^3$He.

{\it Coherent precession.---}Liquid $^3$He in our sample becomes superfluid at $0.95T_c$, where $T_c$ is the superfluid transition temperature in bulk $^3$He. In the temperature range of our measurements, down to $0.3T_c$, only the polar phase is observed. In the polar phase the NMR frequency is given by \cite{AI}
\begin{equation}
\label{spectrum}
\omega=\omega_L + \frac{\Omega^2_P}{2\omega_L}\left[\cos\beta-\frac{\sin^2\lambda}{4}\left(5\cos\beta-1\right)\right].
\end{equation}
Here $\beta$ is the deflection angle of the magnetization from the magnetic field direction (Fig.~\ref{BEC}), $\Omega_P$ is the Leggett frequency in the polar phase, $\omega_L = \gamma H$ is the Larmor frequency, and $\gamma=2.04\cdot10^8$\,s$^{-1}$\,T$^{-1}$ is the absolute value of the gyromagnetic ratio of $^3$He. Most of our experiments are performed in transverse magnetic field ($\lambda=90^\circ$). In this case in cw NMR, where $\cos\beta\approx1$, the frequency shift $\Delta\omega=\omega-\omega_L$ equals zero.

Coherent precession of magnetization is stable only if $d\omega/d(\cos\beta) < 0$ \cite{Fom2}, which corresponds to repulsion between magnons, $d\mu/dn_M>0$. Here the magnon density $n_M = (S-S_z)/\hbar=(\chi H/\gamma\hbar)(1-\cos\beta)$ and $\chi$ is the magnetic susceptibility. In the polar phase the stability condition is satisfied when $|\tan\lambda|>2$, while the tipping angle of magnetization $\beta$ can be arbitrary. The critical magnetic field direction $\lambda_{\rm c} = \arctan 2 \approx 63.4^\circ$. In the stable region superfluid spin currents act to maintain the coherent precession by redistributing magnetization (and $n_M$) across the sample in such a way that the precession frequency $\omega$ in Eq.~(\ref{spectrum}) remains uniform even if $\omega_L$, $\lambda$ and $\Omega_P$ have spatial dependence due to field inhomogeneity and disorder in nafen.

The coherent precession is observed in cw NMR experiment as follows: We initially apply magnetic field $H > \omega_{\rm rf}/\gamma$, where $\omega_{\rm rf}$ is a fixed frequency of rf excitation. Then we gradually decrease $H$. While resonance condition is approached, magnetization deflects and $\beta$ increases, which results in a positive frequency shift of precession $\Delta\omega>0$ according to Eq.~(\ref{spectrum}). When $\omega_L$ becomes smaller than $\omega_{\rm rf}$ during the field sweep, this frequency shift may compensate the difference, and $\omega$ in Eq.~(\ref{spectrum}) becomes locked to $\omega_{\rm rf}$ despite the fact that $\omega_L$ is changing. For this locking to occur, the rf excitation should be large enough to compensate the magnetic relaxation which is presumably determined by the large surface area of the nafen sample. An example of the NMR signals measured in this way is shown in Fig.~\ref{BEC}(a). As one can see in Fig.~\ref{BEC}(b), $\bf M$ can be deflected by more than $90^\circ$. The dissipation grows with increasing $\beta$ and eventually the precessing state collapses, in this case at $\beta\approx130^\circ$.

The coherent nature of the created state is revealed during its decay.
After switching off the rf pumping, magnetic relaxation results in a gradual decrease of $N_M$ and of the amplitude of precession $\sqrt{M_x^2+M_y^2} \propto 1-\cos\beta$. Simultaneously the frequency of precession $\omega$ changes in such a way that relation of Eq.~(\ref{spectrum}) remains valid. This is only possible if the precession remains coherent during the decay and dephasing owing to the magnetic field inhomogeneity $\Delta H / H \approx 8\cdot 10^{-4}$ does not occur. In the absence of dephasing the decay rate in Fig.~\ref{lines}(a) is a measure of the energy relaxation, just as the absorption $M_y$ in cw NMR spectrum, $\dot{E}=\gamma H_{\rm rf}HM_{y}$. Indeed, pulsed and cw measurements of dissipation agree within 30\%, providing further evidence for the coherent precession during the decay.

The structure of the coherently precessing state and the resulting energy dissipation $\dot E$ is essentially given by the distribution of $\beta$ over the sample. The distribution has only a weak dependence on $H_{\rm rf}$ at fixed $\Delta\omega$. Therefore, the absorption signal ($\propto M_y$) is approximately proportional to $H^{-1}_{\rm rf}$, as seen in Fig.~\ref{lines}(b). In the temperature range $(0.38\div0.62)\,T_c$, where we are able to measure dissipation in cw NMR at a fixed tipping angle $\beta=90^\circ$, we have found that the dissipation slightly increases (by about 15\%) on warming. At higher temperatures magnon BEC is destroyed before reaching this value of $\beta$.

{\it Nambu-Goldstone mode.---}The spin rigidity of magnon BEC allows for relatively low-frequency oscillations of the magnetization on the background of the coherent precession. This oscillating mode has a relativistic spectrum
\begin{equation}
\Omega^2 = \mathcal{M}^2 + c^2 k^2,
\label{ngspec}
\end{equation}
where $\Omega$ is the frequency, $k$ is the wave vector of the oscillations and $c$ is the propagation velocity. For a pure NG mode resulting from spontaneous $U(1)$ symmetry breaking in magnon BEC, the mass (or gap) $\mathcal M$ is zero. If magnon BEC is supported by pumping, like in our cw NMR experiments, then explicit breaking of $U(1)$ symmetry by rf field opens gap in the spectrum, and the mode becomes pseudo-Nambu-Goldtsone. In the polar phase this gap is given by \cite{metric}
\begin{equation}
\label{mass}
\mathcal{M}^2=\frac{\Omega^2_P}{8}\frac{H_{\rm rf}}{H}\left(1-5\cos^2\lambda\right)\sin\beta\cos\alpha,
\end{equation}
where the factor $\cos\alpha$ accounts for the fact that oscillations of the phase of precession occur around non-zero $\alpha$ owing to the dissipation.

Boundary conditions in our sample (vanishing spin current through the boundary) allow for spatially uniform oscillations with $k=0$ and frequency $\Omega\equiv\mathcal M$. In the experiment this mode is excited with an alternating field gradient along $\bf H$. Oscillations of $\alpha$ result in periodic variation of the NMR signal. The absorption/dispersion signal is detected by a lock-in amplifier at the frequency $\omega$ and the output is wired to the input of a second lock-in tuned to the frequency of the gradient modulation. Usign the second lock-in we record secondary absorption and dispersion signals as a function of the modulation frequency, as illustrated in the inset in Fig.~\ref{mode}(a). The main peak is fitted by a Lorentzian to obtain the resonance frequency of the pseudo-NG mode $\mathcal{M}$. The secondary spectrum also shows other peaks probably corresponding to standing waves of a pseudo-NG mode with finite $k$, but a detailed study of that is beyond the scope of the present work.

\begin{figure}[t]
\includegraphics[width=\columnwidth]{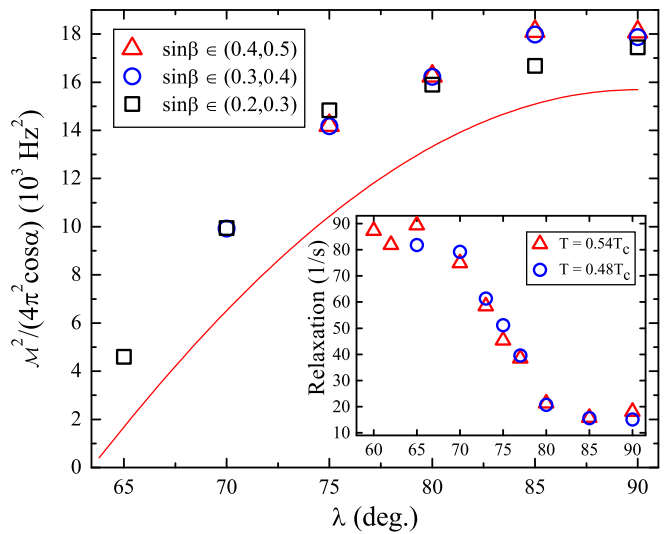}
\caption{(color online).
The  mass $\mathcal{M}$ of the pseudo-NG mode as a function of the magnetic field orientation $\lambda$. Symbols represent ranges of $\sin\beta$ over which the mass has been averaged. The curve is the theoretical dependence for $\sin\beta=0.45$ according to Eq.~(\ref{mass}). For smaller $\beta$ theoretical line goes lower.  The measurements have been done with $H_{\rm rf}$ increasing from 0.16\,$\mu$T to 0.48\,$\mu$T as $\lambda$ decreases from $90^\circ$ to $65^\circ$, but are scaled in the plot with Eq.~(\ref{mass}) to $H_{\rm rf} = 0.71\,\mu$T, which coincides with the data in Fig.~\ref{mode}(c). Temperature is within the range between 0.44$T_c$ and 0.49$T_c$. (Inset) The relaxation rate of magnon BEC as a function of $\lambda$ at different temperatures.}
\label{lambdadep}
\end{figure}

The pseudo-NG mass $\mathcal M$ is plotted in Fig.~\ref{mode} as a function of $H_{\rm rf}$, $\Omega_P$ (controlled by temperature $T$), and $\Delta\omega$, and in  Fig.~\ref{lambdadep} as a function of $\lambda$. Discrepancies at small $\beta$ in Fig.~\ref{mode}(c) and Fig.~\ref{lambdadep} probably originate from $\Delta\omega$ being comparable with the cw NMR linewidth ($\approx300$\,Hz). However, for $\sin\beta > 0.4$  the experimental results are in decent agreement with the theory given by Eq.~(\ref{mass}).
 
The relaxation rate $\tau^{-1}$ of magnon BEC as a function of the field orientation $\lambda$ is shown in the inset of Fig.~\ref{lambdadep}. It is measured by pulsed NMR, and the amplitude of the free induction decay signal is fitted by $\exp(-t/\tau)$. Compared to cw NMR measurements in Fig.~\ref{BEC}, typical  $\beta$ in this measurement is smaller, $\beta \lesssim 10^\circ$, but the precession is still coherent. As expected, the magnon BEC shows maximum stability in the transverse field $\mathbf H \perp \hat{\mathbf m}$. With decreasing $\lambda$ the relaxation rapidly increases and close to the critical angle $\lambda_c$ it is difficult to resolve the coherent precession.

{\it Conclusions.---} We have created a coherently precessing spin state in the polar phase of superfluid $^3$He confined in nafen. The coherent state is observed in cw and pulsed NMR when large enough number of magnons is pumped by rf field. This state has all the signatures of magnon BEC, supported by superfluid spin currents. In particular, its decay in the absence of pumping proceeds only via magnon loss. No dephasing of precession occurs and coherence is preserved by spin supercurrents.
The  broken $U(1)$ symmetry is manifested by the pseudo-Nambu-Goldstone collective mode of coherent precession \cite{Watanabe2013,Nitta2015}. We have measured this mode using resonant excitation and found that its frequency is in close agreement with the theory.

Magnon BEC proved to be an excellent tool to study topological superfluid $^3$He-B. The polar phase opens new possibilities to use magnon BEC as an instrument to probe various topological objects, like half-quantum vortices \cite{HQV}, to manipulate effective metric for NG bosons including modelling black-hole horizon using dependence of $c$ in Eq.~(\ref{ngspec}) on $\lambda$ \cite{metric}, and to investigate the physics of ``relativistic'' fermions living in the vicinity of the Dirac line, where a new type of the quantum electrodynamics emerges \cite{Shou-ChengZhang2017}.

\begin{acknowledgments}
This work has been supported by the Academy of Finland (Projects No.~284594 and No.~298451) and by the European Research Council (ERC) under the European Union's Horizon 2020 research and innovation programme (Grant Agreement No. 694248). We used facilities of the Low Temperature Laboratory infrastructure of Aalto University. Some preliminary experiments have been performed at the Kapitza Institute and were supported by Basic Research Program of the Presidium of Russian Academy of Sciences and by Russian Foundation for Basic Research Grant No. 16-02-00349.
\end{acknowledgments}

\end{document}